%% file: main.tex
\documentclass[conference,a4paper]{IEEEtran}
\IEEEoverridecommandlockouts

\usepackage[utf8]{inputenc} 
\usepackage[T1]{fontenc}
\usepackage{url}              % provides \url{...}
\usepackage[noadjust]{cite}   % improves presentation of citations

\usepackage[cmex10]{amsmath}  % Use the [cmex10] option to ensure complicance
\usepackage{amssymb}          
\usepackage{mleftright}       % fix to wrong spacing of \left-,
\mleftright                   % \middle- \right-commands 

\usepackage{graphicx}         % provides \includegraphics{...} to
\usepackage{booktabs}         % fixes poor spacing in tables and
\newtheorem{theorem}{Theorem}

\newtheorem{definition}[theorem]{Definition}
\newtheorem{corollary}[theorem]{Corollary}
\newtheorem{lemma}[theorem]{Lemma}

\newtheorem{remark}[theorem]{Remark}

\makeatletter
 \renewcommand*\l@subsection{\@dottedtocline{2}{0.3em}{2.2em}}
 \renewcommand*\l@subsubsection{\@dottedtocline{2}{1.2em}{3.2em}}
\makeatother

\begin{document}

\title{The Rate-Distortion-Perception Trade-off\\ with Side Information\\ 
\thanks{The present work has received funding from the European Union’s Horizon 2020 Marie Sk\l{}odowska Curie Innovative Training Network Greenedge (GA. No. 953775). It has also received funding from UKRI (EP/X030806/1) for the project AIR (ERC-CoG). For the purpose of open access, the authors have applied a Creative Commons Attribution (CCBY) license to any Author Accepted Manuscript version arising from this submission.}
} 

%%%%%%

\author{%
  \IEEEauthorblockN{Yassine Hamdi\IEEEauthorrefmark{1}, Deniz G\"{u}nd\"{u}z\IEEEauthorrefmark{1}}
  \IEEEauthorblockA{\IEEEauthorrefmark{1}Department of Electrical and Electronic Engineering, Imperial College London, UK,}
  \IEEEauthorblockA{\{y.hamdi, d.gunduz\}@imperial.ac.uk}
}

\maketitle

%%%%%
%% Abstract: 
%% If your paper is eligible for the student paper award, please add
%% the comment "THIS PAPER IS ELIGIBLE FOR THE STUDENT PAPER
%% AWARD." as a first line in the abstract. 
%% For the final version of the accepted paper, please do not forget
%% to remove this comment!
%%

\input{Other sections/abstract}

\input{Other sections/intro}

\input{Problem formulation and first property/main_file_problem_formulation_and_first_property.tex}

\input{Other sections/section_main_result}

\input{Converse/main_file_converse.tex}

\input{Achievability/main_file_achievability.tex}

\input{Gaussian/main_file_Gaussian.tex}
\enlargethispage{-1.4cm} 
%%%%%%

%%%%%
%%% Do not uncomment for double-blind review submission!
% \section*{Acknowledgments}
%
% We are indebted to Michael Shell for maintaining and improving
% \texttt{IEEEtran.cls}. 

\section{Conclusion}
\label{sec:conclusion}
We have considered the traditional problem of source coding of a memoryless source $X^n$ in the presence of correlated side information $Z^n,$ studied by Wyner and Ziv, with the additional requirement of perfect realism on the reconstruction. We have characterized the rate-distortion-perception trade-off for sources on general alphabets when infinite common randomness is available between the encoder and the decoder, in two cases: when $Z^n$ is available only at the decoder or at both the encoder and the decoder. We showed that, similarly to traditional source coding with side information, the two cases are equivalent when $X^n$ and $Z^n$ are jointly Gaussian. We also provided a general inner bound in the case of limited common randomness.

%%%%%%
%% To balance the columns at the last page of the paper use this
%% command somewhere at the top of the first column of the last page:
%%
% \enlargethispage{-5cm} 
%%
%% where the exact amount of page reduction has to be adapted to the
%% actual situation.
%%
%% If the balancing should occur in the middle of the references, use
%% the following trigger:
%%
% \IEEEtriggeratref{3}
%%
%% which triggers a \newpage (i.e., new column) just before the given
%% reference number. Note that you need to adapt this if you modify
%% the paper. The "triggered" command can be changed if desired:
%%
% \IEEEtriggercmd{\enlargethispage{-20cm}}
%%
%%%%%%

\newpage
\IEEEtriggeratref{15}
\bibliographystyle{IEEEtran}
\bibliography{biblio}

%\input{Appendix/main_file_appendix.tex}

% \newpage
% \tableofcontents

\end{document}

%% file: Other sections/abstract.tex
\begin{abstract}
In image compression, with recent advances in generative modeling, the existence of a trade-off between the rate and the perceptual quality has been brought to light, where the perception is measured by the closeness of the output distribution to the source. This leads to the question: how does a perception constraint impact the trade-off between the rate and traditional distortion constraints, typically quantified by a single-letter distortion measure? We consider the compression of a memoryless source $X$ in the presence of memoryless side information $Z,$ studied by Wyner and Ziv, but elucidate the impact of a perfect realism constraint, which requires the output distribution to match the source distribution. We consider two cases: when $Z$ is available only at the decoder or at both the encoder and the decoder. The rate-distortion trade-off with perfect realism is characterized for sources on general alphabets when infinite common randomness is available between the encoder and the decoder. We show that, similarly to traditional source coding with side information, the two cases are equivalent when $X$ and $Z$ are jointly Gaussian under the squared error distortion measure. We also provide a general inner bound in the case of limited common randomness.
\end{abstract}

%% file: Other sections/intro.tex
\section{Introduction}
\label{sec:intro}

%\todo[inline]{Change title to include perfect realism, because word trade-off is a bit much}

%\todo[inline]{Specify more that fixed-rate, not variable rate}

In conventional rate-distortion theory, the goal is to enable the decoder to reconstruct a representation $Y^n\triangleq (Y_1, ..., Y_n)$ of the source signal $X^n=(X_1, ..., X_n)$ that is close to the latter for some distortion measure $d(X^n,Y^n).$ Shannon characterized the optimal rate-distortion trade-off under an additive distortion measure, i.e. $d(x^n,y^n) = (1/n)\sum_{i=1}^n d(x_i,y_i)$. In \cite{Wyner&Ziv1977}, Wyner and Ziv generalized this result to the case where a side information $Z^n,$ correlated with $X^n,$ is available either only at the decoder or at both the encoder and the decoder.\\
Recently, there has been a renewed interest in compression algorithms since methods based on deep neural networks (DNNs) 
%have received a great deal of attention in recent years. These methods 
have been shown to outperform traditional image and video compression codecs 
\cite{ 2017BalleEndToEndCompression, 2018MentzerDeepImageCompr, ChannelWiseEntropyModelsforLearnedImageCompression2020, KankanahalliEndToEndSpeechCompression2018, KleijnLimWavenetSpeechCoding2018,PetermannBeackHarpNet2021, ZeghidourLuebsSoundStreamEndToEndNeuralAudioCodec2022, RN361, RN363, RN364, RN365, RN367, RN368}
%in image compression 
% \cite{VariableRateImageCompressionRNNs2016, RealTimeAdaptiveCompression2017, 2017BalleEndToEndCompression, JointAutoregressivePriorsForLearnedImageCompression2018, 2018MentzerDeepImageCompr, ChannelWiseEntropyModelsforLearnedImageCompression2020}
% %, speech compression 
% \cite{KankanahalliEndToEndSpeechCompression2018, KleijnLimWavenetSpeechCoding2018}
% %, audio compression 
% \cite{PetermannBeackHarpNet2021, ZeghidourLuebsSoundStreamEndToEndNeuralAudioCodec2022}
%, and video compression 
%\cite{RN360, RN361, RN362, RN363, RN364, RN365, RN366, RN367, RN368} 
under different distortion measures. 
%A main advantage of DNN-based compression methods is their flexibility; they can be trained for different loss functions, in particular directly for the SSIM or MS-SSIM metrics, and they typically lead to remarkable improvements compared to conventional compression algorithms in terms of the average performance. 
%However, it has been shown in \cite{HumanPerceptualEvaluationsNeuralCompression2019} that the reported improvements in the MS-SSIM performance may be misleading for the perceptual quality of the reconstructed images. Furthermore, 
%It has been observed that perceived quality of reconstructed images can be improved using generative adversarial networks (GANs). 
In \cite{2019AgustssonMentzerGANforExtremeCompression}, the authors used generative adversarial networks (GANs) to push the limits of image compression in very low bit-rates by synthesizing image content, such as facades of buildings, using a reference image database. This allows the receiver to generate images that resemble the source image semantically, although they may not match perfectly in details, providing visually pleasing reconstructions even at very low bit-rates. It has been observed (\cite{2011LiEtAlDistributionPreservingQuantization,2018GenerativeCompression,TschannenAgustssonNeurips2018,2019AgustssonMentzerGANforExtremeCompression}) that at such bit-rates the increase in perceptive quality comes at the cost of increased distortion, and the \textit{rate-distortion-perception trade-off} was formalized in \cite{Aug2018MatsumotoRDPDeterministic,Nov2018MatsumotoRDPDeterministic,2019BlauMichaeliRethinkingLossyCompressionTheRDPTradeoff}.
%with respect to the source image.

%\todo[inline]{Check that recently added references appear correctly in bibliography}

%Image perceptual quality has been a motivator, although the origins of the problem dates back longer to Li et al.

%GAN-based compression methods at very low bit-rates.

%It has been shown that there is a trade-off.

Motivated by successful results in generative modeling, where the generated images
%In the extreme case of zero-compression rate, unlike conventional compression codecs which would output a purely deterministic image consisting of average pixel values, a generative model can generate 
would exhibit the same statistical properties of the images in the dataset,
%; hence, with perfect perception quality. As the compression rate increases, we expect the reconstruction to better resemble the particular source image. 
%This idea has been identified as the 
the formalism of
distribution-preserving image compression \cite{2011LiEtAlDistributionPreservingQuantization}
was adopted
in \cite{TschannenAgustssonNeurips2018}, and extended in \cite{Aug2018MatsumotoRDPDeterministic,Nov2018MatsumotoRDPDeterministic,2019BlauMichaeliRethinkingLossyCompressionTheRDPTradeoff}. The problem is then to characterize the optimal rate for which both the distortion $d(X^n, Y^n) \leq \Delta,$ and the perception $\delta(P_{X^n}, P_{Y^n}) \leq \lambda,$
%between the reconstruction distortion and the perception quality. The latter is quantified by a similarity measure 
where
$\delta$ 
%between the source distribution and the reconstruction distribution, 
is a similarity measure,
e.g., the total variation distance or a divergence. This \textit{strong perception constraint} can be replaced by weaker variants \cite{Dec2022WeakAndStrongPerceptionConstraintsAndRandomness}.
In conventional rate-distortion theory, it is known that deterministic encoders are sufficient to achieve the optimal rate-distortion performance. This simplifies both the analysis and implementation of rate-distortion optimal codes. However, for the rate-distortion-perception trade-off, it has been shown in \cite{2021TheisAgustssonOnTheAdvantagesOfStochasticEncoders} that stochastic encoders can be strictly better. This is extensively studied in \cite{Dec2022WeakAndStrongPerceptionConstraintsAndRandomness}.
%than their deterministic counterparts. 
%The authors show that stochastic encoders can be particularly beneficial when near-perfect perfect realism is desired; that is, when $\lambda \to 0.$ This points to the requirement of some common randomness.
%\todo[inline]{Attention, ce que je dis sur la contribution de TheisWagner2021 et Wagner2022 est peut-etre faux du fait de Matsumoto2018}
%\todo[inline]{Include article of Sladi et al, 2015}
%\todo[inline]{Include article of Jun Chen, Dec 2022}

The characterization of the optimal rate-distortion-perception trade-off for variable-rate codes and arbitrary perception measures is given in \cite{TheisWagner2021VariableRateRDP}. See also \cite{Nov2018MatsumotoRDPDeterministic} in the case of a deterministic decoder and for general information sources and \cite{Dec2022WeakAndStrongPerceptionConstraintsAndRandomness} in the case of weaker perception constraints. The characterization for fixed-rate codes for the perfect realism case, i.e., $\lambda=0,$ is given in \cite{2015RDPLimitedCommonRandomnessSaldi} and generalized in \cite{2022AaronWagnerRDPTradeoffTheRoleOfCommonRandomness} to a larger family of distortion measures and alphabets.
%under the total variation distance as the perception measure and 
The impact of the amount of available common randomness on the achievable trade-off is explicitly shown. The optimal rate-distortion trade-off for perfect perception is also explicitly characterized \cite{2022AaronWagnerRDPTradeoffTheRoleOfCommonRandomness} for a Gaussian source and mean-squared error distortion measure. When sufficient common randomness is available, this result boils down to the one in \cite{2011LiEtAlDistributionPreservingQuantization}.

%Wagner and Theis \cite{TheisWagner2021VariableRateRDP}

%Define both words \textit{perception} or \textit{realism}.

%Formalism with distance between probabilities, which comes after the formulation by Li et al with perfect distribution preservation.

First used in another context, a random coding technique to construct distribution-matching stochastic encoder-decoder pairs was developed in \cite{2013PaulCuffDistributedChannelSynthesis} and \cite{2016SongCuffLikelihoodEncoder}. It involves the \textit{soft covering lemma}, and is central in \cite{2022AaronWagnerRDPTradeoffTheRoleOfCommonRandomness} and the present paper. 
%See also  \cite{2014YassaeeOSRB} for a technique in the same line of research.

%\todo[inline]{Remove Yassaee et al from the above paragraph: I am speaking too briefly about strong coordination, so no way to insert Yassaee et al}

%Give credit to \cite{2013PaulCuffDistributedChannelSynthesis} and \cite{2016SongCuffLikelihoodEncoder} for building codes by defining and modifying joint distributions, and analyzing them with the soft covering lemma.

Here, we further push the understanding of how perception constraints affect traditional information-theoretic results by studying the impact of the near-perfect realism constraint, i.e. $\lambda=0,$ on the traditional problem of lossy compression in the presence of side information. We consider two cases: when the side information $Z^n$ is available only at the decoder or at both the encoder and the decoder. In Section \ref{sec:problem_formulation_and_equivalence}, after formalizing the problem, we prove the equivalence between achievability between two notions of asymptotically perfect realism, similarly to \cite{2022AaronWagnerRDPTradeoffTheRoleOfCommonRandomness}. This result, interesting in itself, is key to handling general alphabets in the rest of this paper. In Section \ref{sec:main_result}, we state our main result: a single-letter characterization of the rate-distortion region for fixed-rate codes
%in the two cases 
when infinite common randomness is available between the encoder and the decoder, and an inner bound for limited common randomness. The proof is the subject of Sections \ref{sec:converse} and \ref{sec:achievability}. In Section \ref{sec:gaussian}, we show that, similarly to traditional source coding with side information, the two cases are equivalent when $X$ and $Z$ are jointly Gaussian.

%% file: Problem formulation and first property/main_file_problem_formulation_and_first_property.tex
\section{Problem formulation and a first property}\label{sec:problem_formulation_and_equivalence}
\subsection{Notation}

\noindent Calligraphic letters such as $\mathcal{X}$ denote sets, except in $p^{\mathcal{U}}_{\mathcal{J}},$ which denotes the uniform distribution over alphabet $\mathcal{J}.$ 
%Random variables are denoted using upper case letters such as $X$ and their realizations using lower case letters such as $x.$\\
%For a distribution $P,$ the expression $P_X$ denotes the marginal of variable $X$ while $P(x)$ denotes a real number, probability of the event $X=x.$ Similarly $P_{X|Y=y}$ denotes a distribution over $\mathcal{X}$ and $P_{X|Y=y}(x)$ the corresponding real number.\\
We denote by $[a]$ the set $\{1, ..., \lfloor a \rfloor\}$ and by $x^n$ the finite sequence $(x_1, ..., x_n).$ 
%and use $x_{[n] \backslash k}$ to denote $(x_1, ..., x_{k-1}, x_{k+1}, ..., x_n)$ and $x_{a:b}$ to denote $(x_a, ..., x_b).$ 
%For two sequences $(a_n)_{n \geq 1}$ and $(b_n)_{n \geq 1}$ of positive real numbers, the notation $a_n$ $\substack{\raisebox{-4pt}{$\sim$} \\ \scalebox{0.5}{$n$$\to$$\infty$}}$ $b_n$ means that $\lim_{n \rightarrow \infty} \tfrac{a_n}{b_n} = 1.$\\
We denote by $\|p-q\|_{TV}$ the total variation distance between distributions $p$ and $q.$ We use $I(X;Y)$ to denote the mutual information $X$ and $Y,$ defined for general alphabets as in \cite{Wyner1978GeneralConditionalMutualInformation}.
%(see Appendix  \ref{app:info_theory_for_general_spaces})
%The abreviations a.s. and a.e. stand for \textit{almost surely} and \textit{almost everywhere}.

\subsection{Definitions}

In this section we formulate the information-theoretical problem for general alphabets. To have the existence of conditional distribution given a joint distribution, we assume all alphabets are Polish spaces. This includes discrete spaces and real vector spaces. We omit measure-theoretic justifications where reasonable.
%, see Appendix \ref{app:measure_theory_general} for key basic definitions and results we rely on.

\begin{figure}[t!]
    \centering\includegraphics[width=0.48\textwidth]{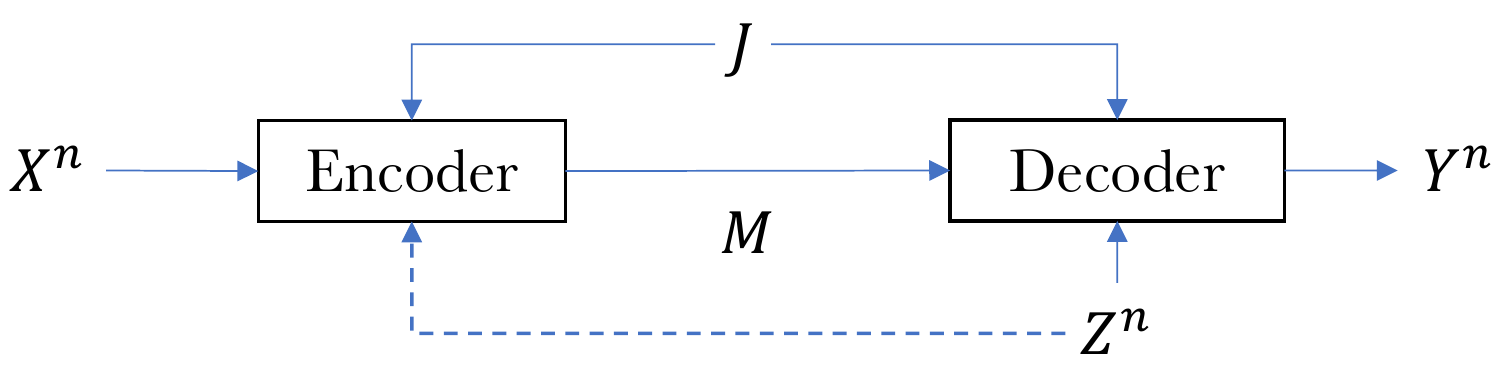}
    \caption{The system model. Side information $Z^n$ is always available at the decoder, but not necessarily at the encoder.}
\label{fig:general_setup}
\end{figure}

\begin{definition}
    Given an alphabet $\mathcal{X},$ a distortion measure is a function $d:\mathcal{X}\times\mathcal{X} \to [0,\infty)$ extending to sequences as $$d(x^n, y^n) = \tfrac{1}{n}\scalebox{0.95}{$\sum_{i=1}^n$} d(x_i,y_i).$$ 
\end{definition}
As shown in Figure \ref{fig:general_setup}, we consider two cases: side information available at the decoder only or available at both the encoder and the decoder.
\begin{definition}
Given two measurable spaces with respective \textit{source} alphabet $\mathcal{X}$ and \textit{side-information} alphabet $\mathcal{Z},$
%and a $\sigma$-finite measure $\mu$ on the former,
a $(n, R, R_c)$ D-code (resp. E-D-code) is a privately randomized encoder and decoder couple $(F^{(n)},G^{(n)})$ consisting of a mapping $F^{(n)}_{M|X^n, J}$ (resp. $F^{(n)}_{M|X^n, Z^n, J}$) from $ \mathcal{X}^n \times [2^{nR_c}] $ (resp. $ \mathcal{X}^n \times \mathcal{Z}^n \times [2^{nR_c}] $) to $ [2^{nR}]$ and a mapping $G^{(n)}_{Y^n|Z^n, J, M}$ from $ \mathcal{Z}^n \times [2^{nR_c}] \times [2^{nR}] $ to $ \mathcal{X}^n.$ 
%dominated by $\mu.$
\end{definition}
\vspace{5pt}
We choose the total variation distance as similarity measure, which allows us to use existing proof techniques for general alphabets, including the soft covering lemma.
\begin{definition}\label{def:achievability}
Given source and side-information alphabets $\mathcal{X}$ and $\mathcal{Z}$ and a joint distribution $p_{X,Z}$ on $\mathcal{X} \times \mathcal{Z},$
%dominated by a product of $\sigma$-finite measures $\mu$ and $\gamma,$ 
a triplet $(R, R_c, \Delta)$ is said to be D-achievable (resp. E-D-achievable) with \textit{near-perfect realism} if there exists a sequence of $(n, R, R_c)$ D-codes (resp. E-D-codes) $(F^{(n)},G^{(n)})_n$ such that 
\begin{IEEEeqnarray}{rCl}
&\limsup_{n \to \infty} \mathbb{E}_{P}[d(X^n,Y^n)] \leq \Delta \quad \text{ and} &\label{eq:def_achievability_distortion_constraint}\\*
&\|P_{Y^n} - p_{X}^{\otimes n}\|_{TV} \underset{n \to \infty}{\longrightarrow} 0, \quad \text{ where}&\label{eq:def_achievability_perception_constraint}\\*
&P_{X^n, Z^n, J, M, Y^n} = p_{X,Z}^{\otimes n} \cdot p^{\mathcal{U}}_{[2^{nR_c}]} \cdot F^{(n)}_{M|X^n, Z^n, J} \cdot G^{(n)}_{Y^|Z^n, J, M}.&\nonumber
\end{IEEEeqnarray} If in addition there exists an integer $N$ such that for all $n\geq N,$ $P_{Y^n} \equiv p_{X}^{\otimes n},$ then $(R, R_c, \Delta)$ is said to be D-achievable (resp. E-D-achievable) with \textit{perfect realism.}
\end{definition}
\vspace{5pt}
The following notion, appearing in \cite{2022AaronWagnerRDPTradeoffTheRoleOfCommonRandomness}, is key to handling general alphabets and is satisfied by finite ones and by the normal distribution with squared-error distortion (Section \ref{sec:gaussian}).
\begin{definition}
    Given a probability space $(\Omega, \mathcal{B}, \mathbb{P}),$ an alphabet $\mathcal{X},$ a probability distribution $p$ on $\mathcal{X}$ and a distortion measure $d,$ we say that $(d,p)$ is uniformly integrable iff for every $\varepsilon >0$ there is a $\tau>0$ such that \begin{equation*}
        \sup_{X,Y,B} \mathbb{E}[d(X,Y)\mathbf{1}_B] \leq \varepsilon,
    \end{equation*} where $X$ and $Y$ represent all variables on $(\Omega, \mathcal{B})$ with law $\mathbb{P}_X \equiv \mathbb{P}_Y \equiv p$ and $B$ represents all events in $\mathcal{B}$ s.t. $\mathbb{P}(B) \leq \tau.$
\end{definition}
\vspace{5pt}

\input{Problem formulation and first property/statement_equivalence.tex}

%\input{Problem formulation and first property/construction_P_2.tex}

%\input{Problem formulation and first property/coupling.tex}

%% file: Problem formulation and first property/statement_equivalence.tex
\subsection{Equivalence of the perfect / near-perfect realism problems}
To prove inner and outer bounds we use (Sections \ref{sec:converse}, \ref{sec:achievability}) the uniform integrability in conjunction with the following results.
\begin{theorem}\label{theorem:equivalence_perfect_realism}
Given Polish source and side-information alphabets $\mathcal{X}$ and $\mathcal{Z},$ a joint distribution $p_{X,Z}$ on $\mathcal{X} \times \mathcal{Z}$
%dominated by a product of $\sigma$-finite measures $\mu$ and $\gamma$ 
and a distortion measure $d$ on $\mathcal{X}$ such that $(d,p_X)$ is uniformly integrable, then a triplet $(R,R_c,\Delta)$ is D-achievable (resp. E-D-achievable) with near-perfect realism if and only if it is D-achievable (resp. E-D-achievable) with perfect realism.
\end{theorem}
\vspace{5pt}
\begin{remark}\label{remark:existence_of_prop_to_perfect_realism}
In Section \ref{sec:achievability}, we use a slightly stronger result: achievability with perfect realism is implied by the existence of codes satisfying \eqref{eq:def_achievability_perception_constraint} and at vanishing total variation distance from a sequence of distributions satisfying \eqref{eq:def_achievability_distortion_constraint}.
\end{remark}

%Proposition \ref{proposition:to_perfect_realism} was proved in 
The proof of Theorem \ref{theorem:equivalence_perfect_realism} and its variant in Remark \ref{remark:existence_of_prop_to_perfect_realism} is the same as that of
\cite{2022AaronWagnerRDPTradeoffTheRoleOfCommonRandomness} 
in the absence of side information.
%in the case $Q\equiv P^{(1)},$ yielding a distribution $P^{(2)}$ satisfying perfect realism and being at negligible total variation distance from $P^{(1)}.$ Using the triangle inequality for the total variation distance produces but a negligible increase and the proof remains valid 
%We provide a proof of the claim of Remark \ref{remark:existence_of_prop_to_perfect_realism} and Theorem \ref{theorem:equivalence_perfect_realism} in Appendix \ref{app:equivalence_perfect_realism}.

%% file: Other sections/section_main_result.tex
\section{Main result}\label{sec:main_result}
Our main result is the full characterization of the region of D-achievable triplets assuming infinite common randomness. We also prove an inner bound for finite common randomness.
\begin{theorem}\label{theorem:D_rates_region}
Consider Polish source and side-information alphabets $\mathcal{X}$ and $\mathcal{Z},$ a joint distribution $p_{X,Z}$ on $\mathcal{X} \times \mathcal{Z}$ 
%dominated by a product of $\sigma$-finite measures $\mu$ and $\gamma$ and 
%such that $I(X;Z) < \infty,$
and a distortion measure $d$ such that $(d,p_X)$ is uniformly integrable. Then, assuming infinite common randomness, the closure of the set $\mathcal{A}_{D,\infty}$ of D-achievable tuples with perfect or near-perfect realism is the closure of the following set $\mathcal{S}_{D,\infty}:$ 
% \begin{IEEEeqnarray}{c}
%     \left\{ \begin{IEEEeqnarray}{rCl} (R, R_c, \Delta) \in \mathbb{R}_{\geq 0}^3 &:& \exists \\ R &\geq& I_p(X;V|Z)\end{IEEEeqnarray}\right\}
% \end{IEEEeqnarray}
\begin{align}\label{eq:def_S_D_infty}
      & 
    \left\{ \begin{array}{rcl}
        (R, \Delta) \in \mathbb{R}_{\geq 0}^2 &:& \exists \ p_{X,Z,V,Y} \in \mathcal{D}_D \text{ s.t. } \\
        R &\geq& I_p(X;V|Z) \\
        \Delta &\geq& \mathbb{E}_p[d(X, Y)]
    \end{array}\right\}, 
\end{align}
where $\mathcal{D}_D$ is defined as
\begin{align}\label{eq:def_D_D}
       & 
    \left\{ \begin{array}{rcl}
        &p_{X,Z,V,Y} : \scalebox{0.9}{$(X,Z) \sim$ } p_{X,Z}, \ p_{Y} \equiv p_{X}&\\
        &Z - X - V, \quad \ X - (Z, V) - Y& \\
        &I_p(Z;V) < \infty&
        %\exists \nu \ \sigma\text{-finite s.t. }& p \ll & \mu \times \gamma \times \nu \times \mu
    \end{array}\right\},
\end{align}where the alphabet of $V$ is constrained to be Polish.
Moreover, in the case of finite common randomness, the closure of the set $\mathcal{A}_D$ of D-achievable triplets $(R,R_c, \Delta)$  with perfect or near-perfect realism contains the closure of the following $\mathcal{S}_D:$
\begin{align}\label{eq:def_S_D}
      & 
    \left\{ \begin{array}{rcl}
        \scalebox{0.9}{$(R, R_c, \Delta) \in \mathbb{R}_{\geq 0}^3$} &:& \scalebox{0.9}{$\exists \ p_{X,Z,V,Y} \in \mathcal{D}_D \text{ s.t. } $}\\
        \scalebox{0.9}{$R$} &\geq& \scalebox{0.9}{$I_p(X;V|Z)$} \\
        \scalebox{0.9}{$R+R_c$} &\geq& \scalebox{0.9}{$I_p(Y;V) - I_p(Z;V)$} \\
        \scalebox{0.9}{$\Delta$} &\geq& \scalebox{0.9}{$\mathbb{E}_p[d(X, Y)]$}
    \end{array}\right\}. 
\end{align}
%The difference $I_p(Y;V) - I_p(Z;V)$ is well-defined by the data processing inequality: $I(Z;V) \leq I(X;V) < \infty.$
% of triplets $(R, R_c, \Delta)$ satisfying:
% \begin{equation}\label{eq:rates_region_R}
%     R \geq I_p(X;V|Z)
% \end{equation}
% \begin{equation}\label{eq:rates_region_R_and_R_c}
%     R+R_c \geq I_p(Y;V) - I_p(Z;V),
% \end{equation}
% \begin{equation}
% \label{eq:rates_region_Markov_distortion}
%     \mathbb{E}_p[d(X^n, Y^n)] \leq \Delta
% \end{equation}
% for some alphabet $\mathcal{V}$ and some transition distributions $(p_{\scalebox{0.6}{$V,Y|X\text{=}x,Z\text{=}z$}})_{(x,z)\in \mathcal{X}\times\mathcal{Z}}$ on $\mathcal{V}\times\mathcal{X}$ belonging to the following set $\mathcal{D}:$
% \begin{equation}\label{eq:rates_region_Markov_short}
%     Z - X - V
% \end{equation}
% \begin{equation}\label{eq:rates_region_Markov_long}
%     X - Z, V - Y^n
% \end{equation}
% \begin{equation}\label{eq:rates_region_perception}
%     p_{Y} \equiv p_{X}
% \end{equation}
% \begin{equation}\label{eq:rates_region_cardinality_bound}
%     |\mathcal{V}| \leq ...
% \end{equation}
\end{theorem}
\vspace{5pt}
%\todo[inline]{Need to prove that from any private randomness on a standard probability space -even discrete- can be replaced by $Z.$ When $Z$ has a continuous cumulative distribution function, then we can find a quantization yielding any discrete law.} 
When the side information $Z$ is independent from the source $X,$ it is independent of $(X,V)$ due to the first Markov chain property. Therefore $I(X;V|Z)=I(X;V)$ and the second Markov property implies $X-V-Y.$ Thus, we recover the result of \cite{TheisWagner2021VariableRateRDP} where no side information is present. 
% because choosing $Z$ to be independent from all variables is in \mathcal{D}_D and does not change the I or E.
%Indeed, given a point from the latter, one can introduce a variable $Z$ independent from all other variables.
%The appearance of conditioning with respect to the side information when the latter is available is reminiscent of the result of Wyner and Ziv \cite{Wyner&Ziv1977} for lossy source coding.\\

%As a corollary of Theorem \ref{theorem:D_rates_region}, in the case of infinite common randomness the region of E-D-achievable couples is also fully characterized.
\vspace{5pt}
\begin{corollary}\label{corollary:E_D_rates_region}
% Consider Polish source and side-information alphabets $\mathcal{X}$ and $\mathcal{Z},$ a joint distribution $p_{X,Z}$ on $\mathcal{X} \times \mathcal{Z}$ 
% %dominated by a product of $\sigma$-finite measures $\mu$ and $\gamma$ and 
% %such that $I(X;Z) < \infty,$ 
% and a distortion measure $d$ such that $(d,p_X)$ is uniformly integrable. Then, assuming infinite common randomness, the closure of the set $\mathcal{A}_{E\text{-}D,\infty}$ of E-D-achievable couples with perfect or near-perfect realism coincides with the closure of the following set $\mathcal{S}_{E\text{-}D,\infty}:$ 
% \begin{align}\label{eq:def_S_E_D}
%       & 
%     \left\{ \begin{array}{rcl}
%         (R, \Delta) \in \mathbb{R}_{\geq 0}^2 &:& \exists \ p_{X,Z,V,Y} \in \mathcal{D}_{E\text{-}D} \text{ s.t. } \\
%         R &\geq& I_p(X;V|Z) \\
%         \Delta &\geq& \mathbb{E}_p[d(X, Y)]
%     \end{array}\right\}, 
% \end{align}
The same result applies for E-D-achievability,
where $\mathcal{D}_D$ is replaced by $\mathcal{D}_{E\text{-}D},$ defined as
\begin{align}\label{eq:def_D_E_D}
       & 
    \left\{ \begin{array}{rcl}
        &p_{X,Z,V,Y} : \scalebox{0.9}{$(X,Z) \sim$ } p_{X,Z}, \ p_{Y} \equiv p_{X}& \\
        &X - (Z,V) - Y& \\
        &I_p(Z;V) < \infty&
        %\exists \nu \ \sigma\text{-finite s.t. }& p \ll & \mu \times \gamma \times \mu
    \end{array}\right\}.
\end{align}
\end{corollary}
\begin{remark}\label{remark:finite_mutual_info}
The region of \cite{2022AaronWagnerRDPTradeoffTheRoleOfCommonRandomness}, taken with $R_c = \infty,$ includes the translation of $\mathcal{S}_{E\text{-}D,\infty}$ by a rate $+I(X;Z).$

% Due to the data processing inequality, this region could be further simplified by setting $V=Y$ if we can guarantee that $I_p(Z;V)<\infty$ for all distributions satisfying the other conditions of \eqref{eq:def_D_E_D}.
% %which is the case if either of $p_X,p_Z$ is discrete or has a well-defined and finite entropy. 
% In that case, $\mathcal{S}_{E\text{-}D,\infty}$ is a translation of the region of \cite{2011LiEtAlDistributionPreservingQuantization},\cite{TheisWagner2021VariableRateRDP} -with $(Z,V)$ considered as a single auxilary variable- by rate $I(X;Z).$
\end{remark}

%We start by showing how Theorem \ref{theorem:D_rates_region} implies this corollary.
Corollary \ref{corollary:E_D_rates_region} follows from applying Theorem \ref{theorem:D_rates_region} with source $(X^n,Z^n)$ instead of $X^n$ and choosing a suitable distortion measure.\\ 
%See Appendix \ref{app:proof_corollary} for details.\\
We prove Theorem \ref{theorem:D_rates_region} in Sections \ref{sec:converse} and \ref{sec:achievability}.

%% file: Converse/main_file_converse.tex
\section{Converse}\label{sec:converse}
We prove that $\overline{\mathcal{A}}_{D,\infty} \subset \overline{\mathcal{S}}_{D,\infty}$ by proving that $\mathcal{A}_{D,\infty} \subset \overline{\mathcal{S}}_{D,\infty}.$ This converse proof builds on the approach of \cite{2012YassaeeEtAlInteractiveCommunicationsForStrongCoordination}, where the same Markov chains as in \eqref{eq:def_D_D} are studied. Let $(R,\Delta)$ be D-achievable with near-perfect realism with infinite common randomness. Then, by Theorem \ref{theorem:equivalence_perfect_realism}, it is D-achievable with perfect realism. Fix $\varepsilon$\hspace{2pt}$>$\hspace{2pt}$0.$
Then there exists a $(n,R, \infty)$ code inducing a joint distribution $P$ such that
$\mathbb{E}_P[d(X^n, Y^n)] \, {<} \, \Delta \, {+} \, \varepsilon$ and $P_{Y^n} \, {\equiv} \, p_X^{\otimes n}.$
% \begin{IEEEeqnarray}{rCl}
% \IEEEeqnarraymulticol{3}{l}{
% \mathbb{E}_P[d(X^n, Y^n)] \leq \Delta + \varepsilon 
% } \label{eq:converse_setting_distortion}\\*
% \text{ and } P_{Y^n} \equiv p_X^{\otimes n}.&& \label{eq:converse_setting_perception}
% \end{IEEEeqnarray}
Let $T$ be a uniform random variable over $[n].$ Define $V\text{=}(M, J, X_{T+1:n}, Z_{1:T-1}, T).$ 
%We shall prove that 
% the single letter distribution $p_{X_T, Y_T, Z_T, V}$ satisfies
% \begin{eqnarray}
%         p_{X_T, Y_T, Z_T, V} &\in& \mathcal{D}_D \nonumber\\*
%         R &\geq& I_p(X_T;V|Z_T) \nonumber\\*
%         \Delta &\geq& \mathbb{E}_p[d(X_T, Y_T)] - \varepsilon. \nonumber
% \end{eqnarray} This will imply that 
%$(R, \Delta+\varepsilon) \in \mathcal{S}_{D,\infty}.$\\
For an i.i.d. vector such as $(X^n,Z^n) \sim p_{X,Z}^{\otimes n},$ the distribution of $(X_T,Z_T)$ is $p_{X,Z}$ 
%because it is a mixture of those of the $(X_i, Z_i).$
Similarly, 
%by \eqref{eq:converse_setting_perception} 
we have $p_{Y_T} \equiv p_X.$
Markov chains in $\mathcal{D}_D$ hold as proved in \cite{2012YassaeeEtAlInteractiveCommunicationsForStrongCoordination}. 
%(see precise reference and remarks on general alphabets in the Appendix \ref{app:Markov_chains_precise_reference_for_converse}). 
%A proof can be found in the Appendix, Section \ref{appendix:simple_justifications}. 
Moreover, we have
\begin{IEEEeqnarray}{rCl}
    % T indep (X_T, Z_T)
    \scalebox{0.85}{$I(V ; Z_T)$} &=&\scalebox{0.85}{$I(M, J, X_{T+1:n}, Z_{1:T-1} ; Z_T | T) $}\nonumber \\*
    &=& \tfrac{1}{n} \scalebox{0.85}{$\sum_{t=1}^n I(M, J, X_{t+1:n}, Z_{1:t-1} ; Z_t) $}\nonumber \\*
    &=& \tfrac{1}{n} \scalebox{0.85}{$\sum_{t=1}^n I(M ; Z_t | J, X_{t+1:n}, Z_{1:t-1}) < \infty,$}\nonumber
\end{IEEEeqnarray} where the first two equalities use the independence of $T$ from all other variables and are true for discrete alphabets. A quantization argument 
%(
based on \cite{Wyner1978GeneralConditionalMutualInformation}
%, see Appendix \ref{app:info_theory_for_general_spaces})
yields the same result for general alphabets. Since the product of Polish spaces is Polish, the alphabet of $V$ is. Thus $p_{X_T, Y_T, Z_T, V} \in \mathcal{D}_D.$
% \noindent We start from \eqref{eq:converse_setting_distortion} and get
% \begin{IEEEeqnarray}{rCl}
% \Delta &\geq& \mathbb{E}[d(X^n, Y^n)] -\varepsilon \nonumber \\*
% &=& \sum_{t=1}^n \dfrac{1}{n}  \mathbb{E}[d(X_t, Y_t)]  -\varepsilon \nonumber \\*
% % &=&  -\varepsilon + \sum_{t=1}^n \dfrac{1}{n} \sum_{(x,y) \in \mathcal{X}^2} P(X_t\text{=}x, Y_t\text{=}y) d(x,y)
% % \nonumber \\*
% % &=&  -\varepsilon + \sum_{t=1}^n P(T\text{=}t) \sum_{\substack{(x,y) \\ \in \mathcal{X}^2}} P(X_t\text{=}x, Y_t\text{=}y|T\text{=}t) d(x,y) \IEEEeqnarraynumspace \label{eq:converse_bounding_distortion} \\*
% % &=&  -\varepsilon + \sum_{t=1}^n \sum_{(x,y) \in \mathcal{X}^2} P(X_T\text{=}x, Y_T\text{=}y, T\text{=}t) d(x,y) \nonumber
% % \\*
% % &=&  -\varepsilon + \sum_{(x,y) \in \mathcal{X}^2} P(X_T\text{=}x, Y_T\text{=}y) d(x,y) \nonumber \\*
% &=& \sum_{t=1}^n \mathbb{E}[\mathbf{1}_{T=t}] \mathbb{E}[d(X_t, Y_t)]  -\varepsilon \nonumber \\*
% &=& \sum_{t=1}^n \mathbb{E}[\mathbf{1}_{T=t} d(X_t, Y_t)]  -\varepsilon \label{eq:converse_bounding_distortion}\\*
% &=& \mathbb{E}[d(X_T, Y_T)] -\varepsilon \nonumber
% \end{IEEEeqnarray} where in \eqref{eq:converse_bounding_distortion} we have use the independence of $T$ from all other variables.
\noindent Using the independence of $T$ from all other variables, we have
\begin{IEEEeqnarray}{c}
    \mathbb{E}[\scalebox{0.85}{$d(X_T, Y_T)$}] = \sum_{t=1}^n \mathbb{E}[\mathbf{1}_{\scalebox{0.6}{$T\text{=}t$}} \scalebox{0.85}{$d(X_t, Y_t)$}] = \sum_{t=1}^n 
    %\mathbb{E}[\mathbf{1}_{\scalebox{0.6}{$T\text{=}t$}}] 
    \mathbb{P}_{\scalebox{0.6}{$T$}}(t)
    \mathbb{E}[\scalebox{0.85}{$d(X_t, Y_t)$}].\nonumber
\end{IEEEeqnarray} Therefore, 
%by \eqref{eq:converse_setting_distortion} 
we have $\Delta + \varepsilon \geq \mathbb{E}[d(X_T, Y_T)].$ Moreover,
\begin{IEEEeqnarray}{rCl}
    \scalebox{0.85}{$nR \geq H(M)$}
%    &\geq& H(M|Z^n, J)\nonumber \\*
    &\geq& \scalebox{0.85}{$I(M;X^n|Z^n, J)$} \nonumber \\
    % J indep of (X^n,Z^n)
    &=& \scalebox{0.85}{$I(M, J ;X^n|Z^n)$} \label{eq:converse_R_using_J_indep} \\
    %just chain rule
    &=& \scalebox{0.85}{$\sum_{t=1}^n I(M, J ; X_t | Z^n, X_{t+1:n})$} \nonumber \\ 
    % (X_{t+1:n}, Z_{[n] \textbactslash t}) indep of (X_t,Z_t)
    &=& \scalebox{0.85}{$\sum_{t=1}^n I(M, J, X_{t+1:n}, Z_{[n] \backslash t} ; X_t | Z_t)$} \nonumber \\
    &\geq& \scalebox{0.85}{$\sum_{t=1}^n I(M, J, X_{t+1:n}, Z_{1:t-1} ; X_t | Z_t)$} \nonumber \\
    &=& \scalebox{0.85}{$n I(M, J, X_{T+1:n}, Z_{1:T-1} ; X_T | Z_T, T)$}\IEEEeqnarraynumspace\label{eq:converse_lower_bound_R_t_to_T} \\  
    % T indep (X_T, Z_T)
%    &=& n I(M, J, X_{T+1:n}, Z_{1:T-1}, T ; X_T | Z_T) \label{eq:converse_lower_bound_R_T_indep} \\*  
    % just definition of V
    &=& \scalebox{0.85}{$n I(V ; X_T | Z_T),$}\label{eq:converse_lower_bound_R_T_indep}
\end{IEEEeqnarray} where \eqref{eq:converse_R_using_J_indep} follows from the independence between the common randomness and the sources and equations \eqref{eq:converse_lower_bound_R_t_to_T} and \eqref{eq:converse_lower_bound_R_T_indep} hold similarly to the above computation of $I(V;Z_T).$
Hence $(R, \Delta+\varepsilon) \in \mathcal{S}_{D,\infty},$ which concludes the proof.

%% file: Achievability/main_file_achievability.tex
\section{Achievability}\label{sec:achievability}

\noindent

\subsection{Informal outline}
We introduce a virtual message $M'$ with rate $R'$ generated and used by the encoder, but not be transmitted. The decoder will then guess it.
We start with a distribution $Q^{(1)},$ analyzed using the soft covering lemma of \cite{2013PaulCuffDistributedChannelSynthesis}, then change it little by little to obtain intermediate distribution $Q^{(2)}$ and a distribution $P^{(1)}$ corresponding to a coding scheme. In order to obtain a final distortion bound, we use the uniform integrability in conjunction with the equivalence between achievability with near-perfect and perfect realism.
For joint distributions, we use, with abuse of notation, $Q_{X,Y}(x,y)=Q_X(x)\rho(y|x),$ which defines $Q_{X,Y}$ by marginal distribution $Q_X$ and conditional probability kernel $\rho.$ 
%See Section \ref{app:measure_theory_general} for more details.

%%%%%%%%%%%%%%%%%%%%%%%%%%%%%%%%%%%%%%%%%%%%%%%%%%%%%%%

\input{Achievability/Codebook.tex}

%%%%%%%%%%%%%%%%%%%%%%%%%%%%%%%%%%%%%%%%%%%%%%%%%%%%%%%%%%%%%%%%

\input{Achievability/Q_1/main_file_Q_1.tex}

%%%%%%%%%%%%%%%%%%%%%%%%%%%%%%%%%%%%%%%%%%%%%%%%%%%%%%%%%%%%%%%%

\input{Achievability/Q_2 and P_1_2/main_file_Q_2.tex}

%%%%%%%%%%%%%%%%%%%%%%%%%%%%%%%%%%%%%%%%%%%%%%%%%%%%%%%%%%%%%%%%

% \input{Achievability/P/main_file_P.tex}

%% file: Achievability/Codebook.tex
\subsection{Random codebook indexed by a virtual message}\label{sec:where_we_fix_epsilon}
Here, we prove that $\overline{\mathcal{S
}}_D \subset \overline{\mathcal{A}}_D$ by proving that $\mathcal{S}_D \subset \overline{\mathcal{A}}_D.$ This will also yield that $\mathcal{S
}_{D, \infty}$ is contained in the projection -along the two coordinates $R$ and $\Delta$- of $\overline{\mathcal{A}}_D,$ which is contained in $\overline{\mathcal{A}}_{D,\infty}.$
Let $(R,R_c,\Delta)$ be a triplet in $\mathcal{S}_D.$ Fix some $\varepsilon >0.$ Let $p_{X,Y,Z,V}$ be a corresponding distribution from the definition of $\mathcal{S}_D.$ Then, we have $I_p(Z;V) < \infty,$ and
\begin{equation}\label{eq:introducing_R_with_eps}
    R \geq I_p(X;V|Z) 
\end{equation}
\begin{equation}\label{eq:introducing_R_c_with_eps}
    R+R_c \geq I_p(Y;V) - I_p(Z;V)
\end{equation}
\begin{equation}\label{eq:introducing_Delta_with_eps}
    \Delta \geq \mathbb{E}_p[d(X,Y)].
\end{equation}
%By the chain rule we have $I_p(X;V|Z) = I_p(X,Z;V) - I_p(Z;V) = (I_p(X;V) - I_p(Z;V|X)) - I_p(Z;V)$ and by the Markov relation $Z-X-V$ in \eqref{eq:def_D_D} this becomes:
By \eqref{eq:def_D_D}, we have
\begin{equation}\label{eq:mutual_info_with_Markov}
    I_p(X;V|Z) = I_p(X;V) - I_p(Z;V).
\end{equation}
% By the chain rule we also have
% \begin{equation}\label{eq:mutual_info_without_any_Markov}
%     I_p(Y;V|Z) = I_p(Y,Z;V) - I_p(Z;V).
% \end{equation} 
We introduce a rate $R'$ corresponding to a virtual message $M',$ as follows. If $I_p(Z;V)=0$ then set $R'=0.$ Otherwise, fix $R'$ in $ \big(I_p(Z;V) - \varepsilon, I_p(Z;V)\big) \cap \big(0,+\infty \big),$ which is possible since $I(Z;V)<\infty.$ Then, by \eqref{eq:introducing_R_with_eps}, \eqref{eq:mutual_info_with_Markov}, \eqref{eq:introducing_R_c_with_eps}, we have:
\begin{IEEEeqnarray}{c}\label{eq:sum_rate_large_enough_for_X_and_Z}
    (R+\varepsilon) + R' > I_p(X;V) = I_p(X,Z;V),\\
\label{eq:sum_rate_large_enough_for_Y}
    (R+\varepsilon) + R' + R_c > I_p(Y;V).
\end{IEEEeqnarray}
For every $n \geq 1$ we generate a random codebook $\mathcal{C}^{(n)}$ with $\lfloor 2^{n(R+\varepsilon)}\rfloor \times \lfloor 2^{nR'}\rfloor \times \lfloor 2^{nR_c}\rfloor$ i.i.d. codewords sampled from $p_V^{\otimes n}.$ The codewords are indexed by triples $(m,m',j).$ We denote this random codebook distribution by $\mathbb{Q}_{\mathcal{C}^{(n)}}.$

%% file: Achievability/Q_1/main_file_Q_1.tex
\subsection{Distribution $Q^{(1)}$ and soft-covering lemma}\label{sec:Q_1}
\begin{figure}[t!]
    \centering\includegraphics[width=0.45\textwidth]{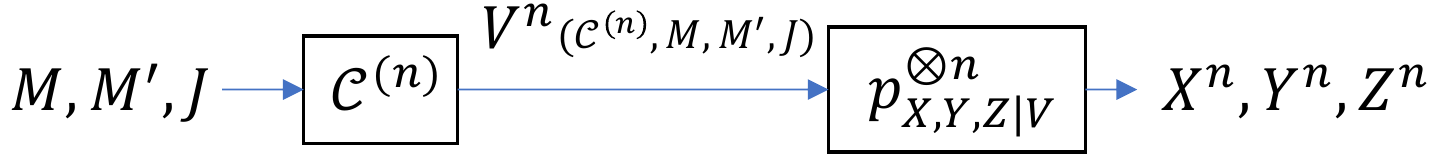}
    \caption{Graphical model for $Q^{(1)}.$}
    \label{fig:soft_covering_setup}
\end{figure}
For every positive integer $n,$ we define the distribution $Q^{(1)},$ described in Figure \ref{fig:soft_covering_setup}, by the following density:
\begin{IEEEeqnarray}{rCl}
\IEEEeqnarraymulticol{3}{l}{
Q^{(1)} \: (c^{(n)},m,m',j,v^n,x^n,y^n,z^n)\ =  \mathbb{Q}_{\mathcal{C}^{(n)}}(c^{(n)})
} \nonumber\\* \quad
&\cdot& \tfrac{\text{\normalsize 1}}{\scalebox{0.7}{$\lfloor 2^{n(R+\varepsilon)} \rfloor \lfloor 2^{nR'} \rfloor \lfloor 2^{nR_c} \rfloor$}} \  \substack{\scalebox{1.0}{$\mathbf{1} \qquad \qquad \qquad \quad$}  \\ v^n = v^n(c^{(n)},m,m',j)} \ \prod_{t=1}^n p \substack{\scalebox{0.8}{$ (x_t, y_t, z_t) $} \\ \scalebox{0.55}{$X, Y, Z|V=v_t$} }\IEEEeqnarraynumspace\label{eq:def_Q_1}
\end{IEEEeqnarray}

%\substack{\scalebox{1.0}{$Q^{(1)} \; (m,m',j,v^n,x^n,y^n,z^n)$} \\ M, M', J, V^n, X^n, Y^n, Z^n \qquad \quad}

%\substack{\scalebox{1.0}{$p^{\mathcal{U}} \; (m,m',j)$} \; \; \\ \scalebox{0.55}{$\lfloor 2^{n(R+\varepsilon)} \rfloor,\lfloor 2^{nR'} \rfloor,\lfloor 2^{nR_c} \rfloor$}}

\noindent Then, by the definition of $\mathbb{Q}_{\mathcal{C}^{(n)}},$ we have $Q^{(1)}_{V^n} \equiv p_V^{\otimes n},$ and therefore, from \eqref{eq:def_Q_1} we have $Q^{(1)}_{X^n, Y^n, Z^n} \equiv p_{X,Y,Z}^{\otimes n}.$ Hence, from \eqref{eq:introducing_Delta_with_eps} and the additivity of $d,$ we have
\begin{equation}\label{eq:Distortion_Q_1}
    \mathbb{E}_{Q^{(1)}}[d(X^n,Y^n)] \leq \Delta.
\end{equation}
\noindent We can use the Markov property $Z-X-V$ to show that 
%$(M,M')$ can be sampled according to $Q^{(1)}$ without relying on $Z^n$ - see the Appendix, Section \ref{appendix:simple_justifications}.
\begin{equation}\label{eq:likelihood_encoder}
    (M,M') - (X^n, J) - Z^n \text{ under } Q^{(1)}.
\end{equation}
%\vspace{5pt}
%%%%%%%%%%%%%%%%%%%%%%%%%%%%%%%%%%%%%%%%%%%%%%%%%%%%%%%%%%%%%%%%%%%%%%%%%%%%

\input{Achievability/Q_1/TV_Q_1.tex}

\input{Achievability/Q_1/Decoding_M_Q_1.tex}

\input{Achievability/Q_1/Choice_of_codebook_for_Q_1.tex}

%% file: Achievability/Q_1/TV_Q_1.tex
\subsubsection{Marginals of $Y^n$ and $(X^n,Z^n)$ knowing a codebook}
\hfill\\
\indent We start by stating a standard lemma regarding the total variation distance. 
%See the Appendix \ref{app:TV_lemmas_for_general_alphabets} for a proof.
%\todo[inline]{Lemma to be typed, probably use the conditional distributions of the common dominating measure}
\begin{lemma}\label{lemma:get_expectation_out_of_TV}
    Let $\Pi$ be two distributions on the product of two Polish spaces $\mathcal{W} $ and $\mathcal{L},$ and let $\Pi_{L|W}, \Gamma_{L|W}$ be two channels. Then, we have \begin{equation*}
        \| \Pi_W \Pi_{L|W} - \Pi_W \Gamma_{L|W} \|_{TV} = \mathbb{E}_{\Pi_W} \big[ \| \Pi_{L|W} - \Gamma_{L|W} \|_{TV} \big].
    \end{equation*}
\end{lemma}
\indent We use the \textit{soft covering lemma} for general alphabets \cite[Corollary~VII.4]{2013PaulCuffDistributedChannelSynthesis} in its memoryless case, and get, as in \cite{2013PaulCuffDistributedChannelSynthesis} 
%-see Appendix \ref{app:use_of_soft_covering} for details-
:
\begin{IEEEeqnarray}{c}
\mathbb{E}_{\mathcal{C}^{(n)}}\big[\|Q^{(1)}_{Y^n|\mathcal{C}^{(n)}} - p_{X}^{\otimes n}\|_{TV}\big] \underset{n \to \infty}{\longrightarrow} 0 \label{eq:TV_Q_1_Y}\\
\mathbb{E}_{\mathcal{C}^{(n)}}\big[\|Q^{(1)}_{J, X^n, Z^n|\mathcal{C}^{(n)}} - p^{\mathcal{U}}_{[2^{nR_c}]}p_{X, Z}^{\otimes n}\|_{TV}\big] \underset{n \to \infty}{\longrightarrow} 0.\IEEEeqnarraynumspace\label{eq:TV_Q_1_J_X_Z}
\end{IEEEeqnarray}

%% file: Achievability/Q_1/Decoding_M_Q_1.tex
\subsubsection{Decoding of $M'$}
\hfill\\
% We define the maximum likelihood decoder corresponding to $p$ as:
% \begin{IEEEeqnarray}{c}
%     \substack{\raisebox{-2pt}{$P^{D}$}\scalebox{1.0}{$\; (m') \qquad \qquad \qquad \qquad \; $} \\ \hat{M}' | \mathcal{C}^{(n)}=c^{(n)}, M=m, J=j, Z^n=z^n} \propto \quad p \substack{\scalebox{1.0}{$(z^n)\qquad \qquad \qquad$} \\ Z^n|V^n=v^n(c^{(n)},m,m',j),}\IEEEeqnarraynumspace
% \end{IEEEeqnarray}
% where the proportionality is with fixed $c^{(n)}, m, j, z^n$ and variable $m'.$ 
\indent The virtual message $M'$ is to be decoded as $\hat{M'},$ conditionally independent from $M',X^n,Y^n$ knowing $M,J,Z^n,\mathcal{C}^{(n)}:$
\begin{equation*}
    \substack{ \raisebox{-2pt}{$Q^{(1)}$} \scalebox{1.0}{$ \; (\hat{m}') \qquad \qquad \qquad \qquad \;$} \\ \hat{M}' | \mathcal{C}^{(n)}=c^{(n)}, M=m, J=j, Z^n=z^n}
\end{equation*}
% \begin{IEEEeqnarray}{rCl}
% \IEEEeqnarraymulticol{3}{l}{
% \substack{\scalebox{1.0}{$Q^{(1)} \; (\hat{m}') \qquad \qquad \qquad \qquad \qquad \qquad \qquad \qquad \qquad \qquad$} \\ \hat{M}' |\mathcal{C}^{(n)}=c^{(n)}, M=m, M'=m', J=j, V^n=v^n, X^n=x^n, Y^n=y^n, Z^n=z^n}
% }\nonumber\\*
% \quad & = & \substack{ \raisebox{-2pt}{$P^{D}$} \scalebox{1.0}{$ \; (\hat{m}') \qquad \qquad \qquad \qquad \;$} \\ \hat{M}' | \mathcal{C}^{(n)}=c^{(n)}, M=m, J=j, Z^n=z^n},\nonumber
%\end{IEEEeqnarray}
%where $P^{D}
is a joint $p_{V,Z}$-typicality decoder for subcodebook $(v^n(c^{(n)},m,a,j))_a.$
By the typical random coding argument for channel coding (see, e.g., \cite{Cover&Thomas2006}), we get
% \noindent On the one hand by definition of $R'$ if $I_p(Z;V)=0$ then $R'=0$ and the alphabet of $M'$ is a singleton, and therefore $M'$ can be decoded perfectly for any choice of $P^{D}.$ On the other hand if $I_p(Z;V)>0$ then $0<R'<I_p(Z;V).$ The codewords in $\mathcal{C}^{(n)}$ are drawn i.i.d. according to $p_V^{\otimes n}$ and, knowing a realization of the codebook, $Q^{(1)}$ behaves according to the memoryless channel with distribution $p_{Z|V}.$ We know (e.g. \cite{Cover&Thomas2006}) that as a consequence, if $\mathcal{Z}$ and $\mathcal{V}$ are finite, with the aforementioned upper bound on $R'$ inducing a bound on the size of every sub-codebook $(v^n(c^{(n)},m,a,j))_a,$ the joint $p$-typicality decoder $P^{D}$ -e.g. from \cite{Cover&Thomas2006}- will be successful:
% \begin{IEEEeqnarray}{c}
% \substack{\scalebox{1.0}{$Q^{(1)} \; (\hat{M}'\neq \ M') $} \\ \hat{M}', M' |M=m, J=j } \underset{n \to \infty}{\longrightarrow} 0. \label{eq:Q_1_decoding_error_fixed_sub_codebook_finite_valued_sources} \IEEEeqnarraynumspace
% \end{IEEEeqnarray}
% Since nothing in this proof depends on the particular value $(m,j),$ including the sub-codebook's distribution, the conditional distribution in \eqref{eq:Q_1_decoding_error_fixed_sub_codebook_finite_valued_sources} does not depend on $(m,j).$ Hence
\begin{IEEEeqnarray}{c}
\scalebox{0.9}{$\substack{\scalebox{1.0}{$Q^{(1)} \; (\hat{M}'\neq M') $} \\ \hat{M}', M' \qquad \qquad} \underset{n \to \infty}{\longrightarrow} 0$}\label{eq:Q_1_decoding_error_fixed_codebook_finite_valued_sources} \IEEEeqnarraynumspace
\end{IEEEeqnarray}
in the case of finite alphabets. 
%See Appendix \ref{app:decoder_M_prime} for details and for a proof 
%of the fact 
It can be proved that the same holds for general sources, using a joint typicality decoder with respect to a quantized distribution $p_{[V], [Z]}$ of mutual information close enough to $I_p(Z;V).$
The probability in \eqref{eq:Q_1_decoding_error_fixed_codebook_finite_valued_sources} can be rewritten as an expectation over $\mathcal{C}^{(n)}.$
% \begin{IEEEeqnarray}{c}
%     \mathbb{E}_{\mathcal{C}^{(n)}}\bigg[ \mathbb{E}_{Q^{(1)}_{M',\hat{M}'|\mathcal{C}^{(n)}=\cdot}} \Big[ \ \substack{\scalebox{1.0}{$\mathbf{1} \qquad $} \\ \scalebox{0.6}{$M' \neq \hat{M}'$}}\Big] \bigg] \underset{n \to \infty}{\longrightarrow} 0.\nonumber
% \end{IEEEeqnarray}
Since the convergence towards zero in expectation implies a convergence in probability for a non-negative variable, we get:
\begin{IEEEeqnarray}{c}\label{eq:decoding_error_Q_1}
     \scalebox{0.9}{$Q^{(1)} \big( M' \neq \hat{M}' \big| \ \mathcal{C}^{(n)} \big) \overset{\mathcal{P}}{\underset{n \to \infty}{\longrightarrow}} 0.$}\IEEEeqnarraynumspace
\end{IEEEeqnarray}

%% file: Achievability/Q_1/Choice_of_codebook_for_Q_1.tex
\subsubsection{Choosing a codebook}
\hfill\\
% The above results imply the existence of a suitable codebook, as follows.
% We start by exhibiting an event on which the average distortion criterion is satisfied.
From \eqref{eq:Distortion_Q_1},
Lemma \ref{lemma:get_expectation_out_of_TV}, and
% we have $$\mathbb{E}_{Q^{(1)}}[d(X^n,Y^n)] \leq \Delta, \text{ i.e. }$$ 
% \begin{IEEEeqnarray}{c}\label{eq:expect_before_Markov_ineq}
%     \mathbb{E}_{\mathcal{C}^{(n)}}\bigg[ \mathbb{E}_{Q^{(1)}_{X^n,Y^n|\mathcal{C}^{(n)}}} [d(X^n,Y^n)] \bigg] \leq \Delta.\IEEEeqnarraynumspace
% \end{IEEEeqnarray}
% For every $n \geq 1$ we define the following deterministic real-valued function: $$ e_n : c^{(n)} \mapsto \mathbb{E}_{Q^{(1)}_{X^n,Y^n|\mathcal{C}^{(n)}=c^{(n)}}} [d(X^n,Y^n)] .$$ Since $e_n \geq 0$ we can use 
the Markov inequality
%, then use \eqref{eq:expect_before_Markov_ineq}:
we get
% \begin{IEEEeqnarray}{c}\label{eq:Markov_ineq_applied}
%     \mathbb{Q}_{\mathcal{C}^{(n)}}(e_n(\mathcal{C}^{(n)}) \leq \Delta + \varepsilon) \geq  1 - \dfrac{\mathbb{E}[e_n(\mathcal{C}^{(n)})]}{\Delta+\varepsilon} \geq \dfrac{\varepsilon}{\Delta+\varepsilon}.\IEEEeqnarraynumspace
% \end{IEEEeqnarray}
\begin{IEEEeqnarray}{c}\label{eq:Markov_ineq_applied}
    \scalebox{0.87}{$\mathbb{Q}_{\mathcal{C}^{(n)}}\Big(\mathbb{E}[d(X^n,Y^n)|\mathcal{C}^{(n)}]\Big) \leq \Delta + \varepsilon) 
    %\geq  1 - \dfrac{\mathbb{E}[e_n(\mathcal{C}^{(n)})]}{\Delta+\varepsilon} 
    \geq \varepsilon/(\Delta+\varepsilon).$}\IEEEeqnarraynumspace
\end{IEEEeqnarray}
In addition, similarly to \eqref{eq:decoding_error_Q_1}, we get convergence in probability from \eqref{eq:TV_Q_1_Y} and \eqref{eq:TV_Q_1_J_X_Z}. Combining this with \eqref{eq:Markov_ineq_applied} gives that for a certain $N_0,$ $\forall n\geq N_0$ there is a codebook $c_*^{(n)}$ such that
\begin{IEEEeqnarray}{c}\label{eq:}
   \scalebox{0.9}{$\mathbb{E}$}_{Q^{(1)}_{X^n,Y^n|\mathcal{C}^{(n)}=c_*^{(n)}}} \scalebox{0.9}{$[d(X^n,Y^n)] \leq \Delta+\varepsilon, $}\label{eq:Distortion_Q_1_fixed_codebook} \IEEEeqnarraynumspace\\
    \scalebox{0.9}{$\|Q^{(1)}_{Y^n|\mathcal{C}^{(n)}=c_*^{(n)}} - p_{X}^{\otimes n}\|_{TV} \leq \varepsilon. $}\label{eq:TV_Q_1_Y_fixed_codebook} \IEEEeqnarraynumspace\\
    \scalebox{0.9}{$\|Q^{(1)}_{J, X^n, Z^n|\mathcal{C}^{(n)}=c_*^{(n)}} - p^{\mathcal{U}}_{[2^{nR_c}]}p_{X, Z}^{\otimes n}\|_{TV} \leq \varepsilon. $}\label{eq:TV_Q_1_J_X_Z_fixed_codebook} \IEEEeqnarraynumspace\\
\scalebox{0.9}{$\substack{\scalebox{1.0}{$Q^{(1)} \; (\hat{M}'\neq M') $} \\ M', \hat{M}' |\mathcal{C}^{(n)}=c_*^{(n)}} \leq \varepsilon. $}\label{eq:Q_1_decoding_error_fixed_codebook} \IEEEeqnarraynumspace
\end{IEEEeqnarray}
In the following sections we shall omit the conditioning on $\mathcal{C}^{(n)}=c_*^{(n)},$ which will be implicit.

%% file: Achievability/Q_2 and P_1_2/main_file_Q_2.tex
\subsection{Construction of a code}
In this subsection, we show how distribution $Q^{(1)}$ can be modified, in a way that is minor in terms of total variation distance, to lead to a code. The latter inherits the realism for $Y^n$ and the small error for decoding $M'.$ Regarding expected distortion, we use the uniform integrability in conjunction with the equivalence between near-perfect and perfect realism.
\vspace{2pt}
\input{Achievability/Q_2 and P_1_2/lemmas_Q_2.tex}

\input{Achievability/Q_2 and P_1_2/Comparison_Q_2_Q_1.tex}

\input{Achievability/Q_2 and P_1_2/P_1_near_perfect_realism.tex}

%% file: Achievability/Q_2 and P_1_2/lemmas_Q_2.tex
\subsubsection{Some lemmas on the total variation distance}
\hfill\\
We start by citing some lemmas from \cite{2013PaulCuffDistributedChannelSynthesis} and \cite{2016SongCuffLikelihoodEncoder}. %See Appendix \ref{app:TV_lemmas_for_general_alphabets} for proofs for general alphabets.
\begin{lemma}\label{lemma:TV_joint_to_TV_marginal}\cite[Lemma~V.1]{2013PaulCuffDistributedChannelSynthesis}
    Let $\Pi$ and $\Gamma$ be two distributions on an alphabet $\mathcal{W} \times \mathcal{L}.$ Then \begin{equation*}
        \| \Pi_W - \Gamma_W \|_{TV} \leq \| \Pi_{W,L} - \Gamma_{W,L} \|_{TV}.
    \end{equation*}
\end{lemma}
\begin{lemma}\label{lemma:TV_same_channel}\cite[Lemma~V.2]{2013PaulCuffDistributedChannelSynthesis}
    Let $\Pi$ and $\Gamma$ be two distributions on an alphabet $\mathcal{W} \times \mathcal{L}.$ Then when using the same channel $\Pi_{L|W}$ we have \begin{equation*}
        \| \Pi_W \Pi_{L|W} - \Gamma_W \Pi_{L|W} \|_{TV} = \| \Pi_W - \Gamma_W \|_{TV}.
    \end{equation*}
\end{lemma}
% \begin{lemma}\label{lemma:TV_continuity}\cite[Expression~(29)]{2013PaulCuffDistributedChannelSynthesis}
%     Let $f$ be a bounded real-valued function and let $\Pi$ and $\Gamma$ be two distributions on an alphabet $\mathcal{W}.$ Then \begin{equation*}
%         \big| E_{\Pi}[f(W)] - E_{\Gamma}[f(W)] \big| \leq 2 f_{\max} \| \Pi - \Gamma \|_{TV}.
%     \end{equation*}
% \end{lemma}
\begin{lemma}\label{lemma:TV_due_to_M_and_hatM}\cite[Lemma~2]{2016SongCuffLikelihoodEncoder}
    Let $P_{UWL}$ be a distribution on an alphabet of the form $\mathcal{U}\times\mathcal{U}\times\mathcal{L}$ and let $\eta \in (0,1).$ If $P(U \neq W) \leq \eta$ we have$\| P_{UL} - P_{WL}\|_{TV} \leq \eta.$
\end{lemma}

%% file: Achievability/Q_2 and P_1_2/Comparison_Q_2_Q_1.tex
\subsubsection{Construction of $Q^{(2)}$ and comparison to $Q^{(1)}$}
\hfill\\
\enlargethispage{-1.1cm}
Using definition \eqref{eq:def_Q_1} of $Q^{(1)}$ and the Markov property $X-Z,V-Y$ from \eqref{eq:def_D_D} we get:
\begin{IEEEeqnarray}{c}
\substack{\scalebox{1.0}{$Q^{(1)} \; (x^n, y^n, z^n)$\quad} \\ X^n, Y^n, Z^n |V^n=v^n} = \prod_{t=1}^n \raisebox{2pt}{\scalebox{1.2}{$p$}} \substack{\scalebox{1.0}{$ (x_t, z_t) $} \\ \scalebox{0.7}{$X, Z|V=v_t$} } \ \raisebox{2pt}{\scalebox{1.2}{$p$}} \substack{\scalebox{1.0}{$ (y_t) \quad \; \; $} \\ \scalebox{0.7}{$Y|V\text{=}v_t, Z\text{=}z_t$} }.\nonumber
\end{IEEEeqnarray}
With this in mind, for every positive integer $n$ we define the following distribution which differs from $Q^{(1)}$ in that $Y^n$ is sampled using $\hat{M}'$ instead of $M':$
\begin{IEEEeqnarray}{rCl}
\IEEEeqnarraymulticol{3}{l}{
Q^{(2)} \: \scalebox{0.9}{$(m,m',j,v^n,x^n,y^n,z^n, \hat{m}')$}
}\nonumber\\* \quad
& = & Q^{(1)} \: \scalebox{0.9}{$(m,m',j,v^n,x^n,z^n, \hat{m}')$} \ \scalebox{1.2}{$p$} \substack{\scalebox{0.95}{$ \; (y^n) \qquad \qquad \qquad \qquad \quad $} \\ \scalebox{0.8}{$Y^n | Z^n\text{=}z^n, V^n\text{=}v^n(m, \hat{m}', j)$} }\nonumber
%\label{eq:def_Q_2}
\end{IEEEeqnarray}
By construction, we have \begin{equation*} Q^{(2)}_{M', \hat{M}', M, J, V^n, X^n, Z^n} \equiv Q^{(1)}_{M', \hat{M}', M, J, V^n, X^n, Z^n}.\end{equation*}
Using Lemma \ref{lemma:TV_due_to_M_and_hatM} on this joint distribution with $U=M', W=\hat{M}',L=(M, J, V^n, X^n, Z^n)$ we have by \eqref{eq:Q_1_decoding_error_fixed_codebook}
\begin{IEEEeqnarray}{c}
    \big\|Q^{(1)}_{M', M, J, V^n, X^n, Z^n} - Q^{(2)}_{\hat{M}', M, J, V^n, X^n, Z^n}\big\|_{TV} \leq \varepsilon. \nonumber
\end{IEEEeqnarray}
As a consequence and by construction of $Q^{(2)}$ and its similarity to that of $Q^{(1)}$ we have by Lemma \ref{lemma:TV_same_channel} with $L=Y^n:$
\begin{IEEEeqnarray}{c}\label{eq:TV_Q_1_Q_2}
    \big\|Q^{(1)}_{\scalebox{0.7}{$M', M, J, V^n, X^n, Y^n, Z^n$}} - Q^{(2)}_{\scalebox{0.7}{$\hat{M}', M, J, V^n, X^n, Y^n, Z^n$}} \big\|_{TV} \leq \varepsilon. \IEEEeqnarraynumspace
\end{IEEEeqnarray}
% \begin{IEEEeqnarray}{rCl}\label{eq:TV_Q_1_Q_2}
%     \big\|&&Q^{(1)}_{M', M, J, V^n, X^n, Y^n, Z^n} \nonumber\\*
%     &\qquad& \qquad - \ Q^{(2)}_{\hat{M}', M, J, V^n, X^n, Y^n, Z^n}\quad \big\|_{TV} 
%     \underset{n \to \infty}{\longrightarrow} 0.\IEEEeqnarraynumspace
% \end{IEEEeqnarray}
% Then by Lemma \ref{lemma:TV_joint_to_TV_marginal} with $W=(X^n, Y^n)$ and by the boundedness of $d,$ Lemma \ref{lemma:TV_continuity} and \eqref{eq:Distortion_Q_1} we have
% \begin{equation}\label{eq:Distortion_Q_2}
%     \limsup_{n \to \infty} \mathbb{E}_{Q^{(2)}}[d(X^n,Y^n)] \leq \Delta + \varepsilon.
% \end{equation}
Since $Q^{(2)}_{J, X^n, Z^n} \equiv Q^{(1)}_{J, X^n, Z^n}$ we also have by \eqref{eq:TV_Q_1_J_X_Z_fixed_codebook}:
\begin{equation}\label{eq:TV_Q_2_J_X_Z}
    \|Q^{(2)}_{J, X^n, Z^n} - p^{\mathcal{U}}_{[2^{nR_c}]}p_{X, Z}^{\otimes n}\|_{TV} \leq \varepsilon.
\end{equation}
% Moreover by Lemma \ref{lemma:TV_joint_to_TV_marginal} applied to \eqref{eq:TV_Q_1_Q_2} with $W=Y^n$ we get  
% \begin{IEEEeqnarray}{c}\label{eq:TV_sur_Y_Q_1_Q_2}
%     \big\|Q^{(1)}_{Y^n} - Q^{(2)}_{Y^n}\big\|_{TV} \leq \varepsilon.
% \end{IEEEeqnarray} 
% By the triangle inequality of the total variation distance and \eqref{eq:TV_Q_1_Y} we get:
% \begin{equation}\label{eq:TV_Q_2_Y}
%     \|Q^{(2)}_{Y^n} - p_{X}^{\otimes n}\|_{TV} \underset{n \to \infty}{\longrightarrow} 0.
% \end{equation}

%% file: Achievability/Q_2 and P_1_2/P_1_near_perfect_realism.tex
\subsubsection{Finalizing the code construction}
\hfill\\
We define the distribution $P^{(1)}$ achieving near-perfect realism and from which a distribution $P^{(2)}$ with perfect realism will be derived. The former differs from $Q^{(2)}$ in having the correct marginal for $(X^n,Z^n)$ as follows
\begin{IEEEeqnarray}{rCl}
\IEEEeqnarraymulticol{3}{l}{
P^{(1)} \: (m,m',j,v^n,x^n,y^n,z^n, \hat{m}')
}\nonumber\\*
%& = &  \nonumber \\*
% &  & \substack{\scalebox{1.0}{$\mathbf{1} \qquad \qquad \qquad$}  \\ v^n = v^n(c^{(n)},m,m',j)} \ \prod_{t=1}^n p \substack{\scalebox{0.8}{$ (x_t, y_t, z_t) $} \\ \scalebox{0.55}{$X, Y, Z|V=v_t$} }\nonumber\\*
&=& \tfrac{\text{\normalsize 1}}{\lfloor 2^{nR_c} \rfloor} \ \prod_{t=1}^n \raisebox{2pt}{\scalebox{1.1}{$p$}} \substack{\scalebox{1.0}{$ \: (x_t, z_t) $} \\ \scalebox{0.7}{$X, Z \qquad \quad$} } \ \substack{\scalebox{1.0}{$Q^{(2)} \; (m,m', v^n, \hat{m}', y^n) \qquad \quad  $} \\ \scalebox{0.7}{$M, M', V^n, \hat{M}', Y^n| J\text{=}j, X^n\text{=}x^n, Z^n\text{=}z^n$}} \nonumber
% &&\substack{ \raisebox{-2pt}{$P^D$} \scalebox{1.0}{$ \; (\hat{m}') \qquad \qquad \qquad \qquad \;$} \\ \hat{M}' | \mathcal{C}^{(n)}=c^{(n)}, M=m, J=j, Z^n=z^n}\nonumber \\*
% &&\substack{\scalebox{1.0}{$Q^{(2)} \; (y^n) \qquad \qquad \qquad \qquad \qquad \qquad \qquad $} \\ \scalebox{0.7}{$Y^n| \mathcal{C}^{(n)}\text{=}c^{(n)}, M\text{=}m, M'\text{=}m', J\text{=}j, X^n\text{=}x^n, Z^n\text{=}z^n, \hat{M}'\text{=}\hat{m}'$}}. \label{eq:def_P_1}
\end{IEEEeqnarray}
% By definitions \eqref{eq:def_Q_1}, \eqref{eq:def_Q_2} of $
% Q^{(1)}$ and $Q^{(2)},$ we have that $P^{(1)}$ a distribution induced be a $(n,R+\varepsilon,R_c)$ D-code.
Then by Lemma \ref{lemma:TV_same_channel} comparing $P^{(1)}$ with $Q^{(2)}$ reduces to comparing marginals, i.e. to equation \eqref{eq:TV_Q_2_J_X_Z}:
\begin{IEEEeqnarray}{rCl}
\IEEEeqnarraymulticol{3}{l}{
\big\|P^{(1)}_{\scalebox{0.6}{$M, M', J, V^n, X^n, Z^n, \hat{M}', Y^n$}} - Q^{(2)}_{\scalebox{0.6}{$M, M', J, V^n, X^n, Z^n, \hat{M}', Y^n$}}\big\|_{TV}
}\nonumber\\*
\quad & = & \big\|P^{(1)}_{J, X^n, Z^n} - Q^{(2)}_{J, X^n, Z^n}\big\|_{TV}\leq \varepsilon. \label{eq:TV_P_1_Q_2}
\end{IEEEeqnarray} 
Therefore by Lemma \ref{lemma:TV_joint_to_TV_marginal} with $W=(X^n, Y^n)$ and the triangle inequality and \eqref{eq:TV_Q_1_Q_2} we get
\begin{IEEEeqnarray}{c}
    \big\|P^{(1)}_{X^n, Y^n} - Q^{(1)}_{X^n, Y^n}\big\|_{TV} \leq \varepsilon. \label{eq:TV_sur_X_Y_P_1_Q_1}
\end{IEEEeqnarray}
Due to \eqref{eq:likelihood_encoder}, it can be easily checked that $P^{(1)}$ defines a $(n,R+\varepsilon,R_c)$ D-code.
The entire construction layed out in this Section \ref{sec:achievability} is valid for any $\varepsilon>0$ (which was fixed in Section \ref{sec:where_we_fix_epsilon}). The blocklength $N_0$ after which \eqref{eq:Distortion_Q_1_fixed_codebook} and \eqref{eq:TV_Q_1_Y_fixed_codebook} hold depends on  $\varepsilon>0.$ Consider a sequence of codes associated to some vanishing sequence $(\varepsilon_k)_k.$
For any $\varepsilon'>0,$ the achievability of $(R+\varepsilon',R_c, \Delta+\varepsilon')$ then follows from Remark \ref{remark:finite_mutual_info} with $Q^{(1)}$ and $P^{(1)}$ and from \eqref{eq:Distortion_Q_1_fixed_codebook}, \eqref{eq:TV_Q_1_Y_fixed_codebook} and \eqref{eq:TV_sur_X_Y_P_1_Q_1}.  
%This proves the achievability of $(R+\varepsilon,R_c, \Delta+\varepsilon + s_{\varepsilon}),$ where $s_{\varepsilon}$ is the supremum from the proposition, which does not depend on $n$ and satisfies $s_{\varepsilon} \to 0$\\
%if $\varepsilon \to 0$ by the uniform integrability assumption. This being true for every $\varepsilon>0,$ 
Hence, $(R, R_c, \Delta) \in \overline{\mathcal{A}}_D$ as desired.

%% file: Gaussian/main_file_Gaussian.tex
\section{The Gaussian case}\label{sec:gaussian}
Interestingly, similarly to standard source coding with side information, when the source and side information form a bi-dimensional Gaussian then D-achievability is equivalent to E-D-achievability in the following sense.
\begin{theorem}\label{theorem:gaussian}
Consider the setting of Theorem \ref{theorem:D_rates_region} in the case of infinite common randomness with
$d:(x,y)\mapsto (x-y)^2$ \begin{equation*}
    \text{and } p_{X,Z} = \mathcal{N}\Big(0, \scalebox{0.8}{$\begin{pmatrix}
1 & \eta \\
\eta & 1
\end{pmatrix}$} \Big).
\end{equation*} For any $\Delta$ in $(0, 2-2\eta],$ and denoting $\rho=1-\Delta/2,$ the infimum of rates such that $(R,\Delta)$ is D- or E-D-achievable with (near-)perfect realism is:
\begin{equation}
    \scalebox{0.93}{$R_{D}(\Delta) = R_{E\text{-}D}(\Delta) = \dfrac{1}{2}\log\Big(\dfrac{1-\eta^2}{1-\rho^2}\Big),$}
\end{equation}
\end{theorem}
The numerator is the same as in standard source coding with side information (see e.g. \cite{Cover&Thomas2006}) and the denominator is the same as in \cite{2011LiEtAlDistributionPreservingQuantization}, which we recover when the side-information $Z$ is independent from $X.$
We now prove Theorem \ref{theorem:gaussian}.\\
% \todo[inline]{Cannot use Remark \ref{remark:finite_mutual_info}}
We know that $R_{D}(\Delta) \geq R_{E\text{-}D}(\Delta).$ Moreover,
%since $p_X,p_Z,p_{X,Z}$ have well-defined and finite entropies, then 
by Remark \ref{remark:finite_mutual_info} the region $\mathcal{S}_{E\text{-}D,\infty}$ translated by a rate $+I(X;Z)=-\tfrac{1}{2}\log((1-\eta^2))$ is included in the region of \cite{2011LiEtAlDistributionPreservingQuantization,TheisWagner2021VariableRateRDP}. Therefore, by \cite[Proposition~2]{2011LiEtAlDistributionPreservingQuantization} we have
\begin{equation}
    \scalebox{0.93}{$R_{E\text{-}D}(\Delta) \geq \tfrac{1}{2}\log((1-\eta^2)/(1-\rho^2)).$}
\end{equation}
We upper bound $R_{D}(\Delta).$
Fix a $\Delta$ in $(0, 2-2\eta].$ Let 
%$\rho=1-\Delta/2$ and
\begin{equation}\label{eq:def_Z_X_V}
    \scalebox{0.9}{$(Z,X,V) = \big(\eta \Tilde{X} + \sqrt{1-\eta^2} \Tilde{Z}, \Tilde{X}, b \Tilde{X} + \sqrt{1-b^2} \Tilde{V}\big),$}
\end{equation}where $(\Tilde{Z}, \Tilde{X}, \Tilde{V})$ is standard Gaussian. 
Since $\rho \geq \eta,$ with
\begin{equation}\label{eq:def_b}
    b= \sqrt{(\rho^2 - \eta^2)/(1+\eta^2\rho^2-2\eta^2)} \text{ we find}
\end{equation}
%For a suitable choice of $b,$ one can find (see Appendix \ref{app:Gaussian}) %conditional means and variances and get
\begin{IEEEeqnarray}{c}
    \scalebox{0.93}{$\mathbb{E}\big[ \mathbb{E}[X|Z,V]^2 \big] = \rho^2$} \text{, and therefore}\label{eq:rho_carre}\\*
    \scalebox{0.90}{$I_p(X;V|Z) = h(X|Z) - h(X|Z,V) = \tfrac{1}{2}\log((1\text{-}\eta^2)/(1\text{-}\rho^2)).$}\label{eq:gaussian_conditional_mutual_info}\IEEEeqnarraynumspace
\end{IEEEeqnarray}
Define $Y=\rho^{-1}\mathbb{E}[X|Z,V].$ 
%This implies that 
One can check that
$p_{X,Z,V,Z} \in \mathcal{D}_{D},$
that
$\mathbb{E}[d(X,Y)]=\Delta$
% with
% \begin{equation}
%     \alpha = \rho^{-1/2} \dfrac{\eta-\eta b^2}{1-\eta^2 b^2} \text{ and } \beta = \rho^{-1/2} \dfrac{b-\eta^2 b}{1-\eta^2 b^2}.
% \end{equation}
and $(d, p_{X})$ is uniformly integrable,
%(see Appendix \ref{app:Gaussian})
 then by Theorem \ref{theorem:D_rates_region},
we have
%equations \eqref{eq:gaussian_conditional_mutual_info} and \eqref{eq:distortion_is_Delta} imply that 
\begin{equation*}
    \scalebox{0.95}{$R_{D}(\Delta) \leq \tfrac{1}{2}\log((1-\eta^2)/(1-\rho^2)).$}
\end{equation*} 